\documentclass[10pt]{article}
\usepackage{times}
\usepackage[usenames]{color}
\def\pd#1#2{\frac{\partial#1}{\partial#2}}
\def\pdd#1#2#3{\ifx#2#3\frac{\partial^2#1}{\partial#2^2}\else\frac{\partial^2#1}{\partial#2\mkern1mu\partial#3}\fi}
\def\dash{\unskip\nobreak\hskip.05555em plus.11111em---\hskip.05555em plus.11111em\relax}

\def\eqnitem#1. {\par\medbreak\leavevmode#1$^\circ$.\enspace}
\def\Example#1.{{\sl Example\hskip.5em\relax#1\unskip.}}
\def\Remark#1.{{\sl Remark\hskip.5em\relax#1\unskip.}}
\let\BL=\biggl \let\BR=\biggr
\let\Bl=\Bigl \let\Br=\Bigr
\let\bl=\bigl 
\let\ds=\displaystyle
\begin{document}

\title{New classes of exact solutions of three-dimensional
Navier--Stokes equations}

\author{%
S.~N.~Aristov\footnote
{Institute of Continuous Media Mechanics, Ural Branch, Russian Academy of Sciences,
  1~Acad.\ Koroleva Str., 614013 Perm, Russia.
  E-mail: asn@icmm.ru}
\and
A.~D.~Polyanin\footnote
{Institute for Problems in Mechanics,
  Russian Academy of Sciences,
  101~Vernadsky Avenue, bldg~1, 119526 Moscow, Russia.
  E-mail: polyanin@ipmnet.ru}}
\maketitle

\begin{abstract}
\noindent \textcolor{blue}{New classes of exact solutions of the
three-dimensional unsteady Navier--Stokes equations containing arbitrary
functions and parameters are described. Various periodic and other solutions,
which are expressed through elementary functions are obtained. The general
physical interpretation and classification of solutions is given.}

\medskip
\textcolor{blue}{{\bf Keywords:} Navier--Stokes equations, exact solutions, periodic solutions,
three-dimensional equations, unsteady equations.}
\end{abstract}

\section{Class of motions of the viscous incompressible fluid under consideration}

\textcolor{blue}{Self-similar, invariant, partially invariant, and certain other exact solutions
of the Navier--Stokes equations including those with generalized separation o
f variables were considered, for example, in [1--15].
Below, the term "exact solutions" is used according to the definition given in
[14, p. 10].}
\medskip

\textcolor{blue}{Three-dimensional unsteady motions of a viscous incompressible
fluid are described by the Navier--Stokes and continuity equations:}
\begin{eqnarray}
\pd {V_1}t+V_1\pd {V_1}x+V_2\pd {V_1}y+V_3\pd {V_1}z&=&-\frac 1{\rho}\pd Px
+\nu\BL(\pdd {V_1}xx+\pdd {V_1}yy+\pdd {V_1}zz\BR),\label{(1)}\\
\pd {V_2}t+V_1\pd {V_2}x+V_2\pd {V_2}y+V_3\pd {V_2}z&=&-\frac 1{\rho}\pd Py
+\nu\BL(\pdd {V_2}xx+\pdd {V_2}yy+\pdd {V_2}zz\BR),\\
\pd {V_3}t+V_1\pd {V_3}x+V_2\pd {V_3}y+V_3\pd {V_3}z&=&-\frac 1{\rho}\pd Pz
+\nu\BL(\pdd {V_3}xx+\pdd {V_3}yy+\pdd {V_3}zz\BR),\\
\pd {V_1}x+\pd {V_2}y+\pd {V_3}z&=&0.
\end{eqnarray}
\textcolor{blue}{Here $x$, $y$, and $z$ are the Cartesian coordinates, $t$ is time;
$V_1$, $V_2$, and $V_3$ are the fuid-velocity components;
$P$~is pressure; and $\rho$ and $\nu$
are the fluid density and kinematic viscosity, respectively.
When writing Eqs.~\hbox{(1)--(4)}, it was assumed that the mass forces are potential
and included in the pressure.}

\textcolor{blue}{We consider the flow of a viscous incompressible fluid when the
fluid-velocity vector on the  $z$ axis is directed along this axis.
Near the $z$ axis, the transverse velocity components are small and
can be expanded in a Taylor series in terms of the transverse $x$ and $y$ coordinates.
If we restrict ourselves to the main terms of the expansion in $x$ and $y$
for the velocity components, it is possible to obtain the following
representation for the desired values after the corresponding analysis:}
\begin{equation}
\begin{array}{rl}
&V_1=x\Bl(-\ds\frac12\pd Fz+w\Br)+yv,\quad V_2=xu-y\Bl(\ds\frac12\pd Fz+w\Br),\quad V_3=F,\\[6pt]
&\quad \ \ds\frac P\rho=p_0-\frac12\alpha x^2-\frac12\beta y^2-\gamma xy-\frac12 F^2
+\nu\pd Fz-\int \pd Ft\,dz,
\end{array}
\label{(5)}
\end{equation}
\textcolor{blue}{where $p_0$, $\alpha$, $\beta$, and $\gamma$ are arbitrary functions of time~$t$
setting the transverse pressure distribution; $F$,~$u$,~$v$, and~$w$
are unknown functions dependent on the coordinate $z$ and~$t$.
The substitution of Eqs.~(5) into Navier--Stokes Eq.~(3) and
continuity Eq.~(4) results in identities, and Eqs.~(1) and (2)
become $A_nx + B_ny = 0$ \ ($n = 1,\,2$),
where $A_n$ and~$B_n$ represent certain differential expressions
dependent on the variables $z$ and~$t$. The split in the variables $x$ and~$y$
results in four equations $A_n = 0$ and $B_n = 0$ \ ($n = 1,\,2$),
which can be transformed to the following form:}
\begin{eqnarray}
\pdd Ftz+F\pdd Fzz-\frac12\Bl(\pd Fz\Br)^2&=&-(\alpha+\beta)+
\nu \frac{\partial^3F}{\partial z^3}+2(uv+w^2),\label{(6)}\\
\pd ut+F\pd uz-u\pd Fz&=&\gamma+\nu\pdd uzz,\\
\pd vt+F\pd vz-v\pd Fz&=&\gamma+\nu\pdd vzz,\\
\pd wt+F\pd wz-w\pd Fz&=&\frac{\alpha-\beta}2+\nu\pdd wzz.
\end{eqnarray}

\textcolor{blue}{It is important to emphasize that solution (5) precisely satisfies Eqs.~\hbox{(1)--(4)}
for the viscous fluid motion by virtue of Eqs.~\hbox{(6)--(9)}.}

\textcolor{blue}{For $\gamma= 0$, the structure of exact solution (5) and system (6)--(9)
was obtained in [12] from other reasons by the investigation of the class of
partially invariant solutions (the case of $\alpha = \beta = \gamma = 0$
was considered in [7]). In [12], the group classification of system (6)--(9)
was carried out for $\gamma = 0$, which resulted in singling out two types of time
dependences for the determining functions: (i)~$\alpha$~and~$\beta$
are constant, and (ii)~$\alpha$~and~$\beta$ are proportional to~$t^{-2}$
(the exact solutions of the Navier--Stokes equations with a reasonably simple
structure correspond to these dependences).}

\textcolor{blue}{In this work, we obtained new classes of exact solutions of system (6)--(9),
when the determining functions contain a functional arbitrariness.
The basic idea of the following analysis is that we can
obtain a single isolated equation for the longitudinal velocity component
$V_3 = F$ from system (6)--(9).}

\section{Reduction of system (6)--(9) to a single equation}

\textcolor{blue}{We consider first the special class of exact solutions described by a single equation.}

\textcolor{blue}{In Eqs.~\hbox{(6)--(9)}, we put}
\begin{equation}
u=m\pd Fz+A,\quad v=n\pd Fz+B,\quad w=k\pd Fz+C,
\label{(10)}
\end{equation}
\textcolor{blue}{where $m$, $n$, $k$, $A$, $B$, and $C$
are the desired functions of time $t$.
We require that four Eqs.~\hbox{(6)--(9)} coincide after the substitution of
Eq.~(10) in them. As a result, for determining the desired functions,
we obtain the nonlinear system consisting of one algebraic equation and
six ordinary differential equations:}
\begin{eqnarray}
mn+k^2&=&\frac14,\label{(11)}\\
\frac{A-m'}m=\frac{B-n'}n&=&\frac{C-k'}k=2(An+Bm+2Ck),\\
\frac{\gamma-A'}m=\frac{\gamma-B'}n&=&\frac{\alpha-\beta-2C'}{2k}=-\alpha-\beta+2AB+2C^2.
\end{eqnarray}
\textcolor{blue}{This system contains seven equations for nine functions\dash six functions
$m$,~$n$, $k$, $A$,~$B$, and~$C$ from Eq.~(10) and three functions
$\alpha$,~$\beta$, and~$\gamma$ from Eqs.~\hbox{(6)--(9)} (in this case, they are
also treated as desired). It is possible to show that the last equation in
Eq.~(12) is the consequence of three other equations (11),~(12).
Therefore, three desired functions in system (11), (13)
can, in general, be chosen arbitrarily.}

\textcolor{blue}{Taking into account Eqs.~\hbox{(10)--(13)}, we reduce system (6)--(9) to a single equation}
\begin{equation}
\pdd Ftz+F\pdd Fzz-\Bl(\pd Fz\Br)^2=
\nu \frac{\partial^3F}{\partial z^3}+q\pd Fz+p,
\label{(14)}
\end{equation}
\textcolor{blue}{the functions $p = p(t)$ and $q = q(t)$ are defined by the relations}
\begin{equation}
p=\frac{\gamma-A'}m,\quad q=\frac{A-m'}m.
\label{(15)}
\end{equation}

\textcolor{blue}{{\it The general property of Eq.~(14)\/}. Suppose $F_0(z,t)$ is a solution of this
equation. Then the function}
\begin{equation}
F=F_0(z+\psi(t),t)-\psi'_t(t),
\label{(16)}
\end{equation}
\textcolor{blue}{where $\psi(t)$ is an arbitrary function, is also a solution of Eq.~(14).}

\textcolor{blue}{For constructing solutions of system (11)--(13), it is necessary to distinguish
two cases.}

\textcolor{blue}{\eqnitem 1.
{\sl Case of $m=n$\/}.
In this case, the general solution of system (11)--(13) can be represented as}
\begin{equation}
\begin{array}{rl}
&m=n=\frac 12\sin\varphi,\quad k=\frac12\cos\varphi,\\[6pt]
&A=B=\frac12 (q\sin\varphi+\varphi'_t\cos\varphi),\quad
C=\frac12 (q\cos\varphi-\varphi'_t\sin\varphi),\\[6pt]
&\alpha=\frac14q^2+\frac14(\varphi'_t)^2-\frac12p(1-\cos\varphi)+C'_t,\\[6pt]
&\beta=\frac14q^2+\frac14(\varphi'_t)^2-\frac12p(1+\cos\varphi)-C'_t,\\[6pt]
&\gamma=\frac12p\sin\varphi+A'_t,
\end{array}
\label{(17)}
\end{equation}
\textcolor{blue}{where $p=p(t)$, $q=q(t)$, and $\varphi=\varphi(t)$ are arbitrary functions.
For convenience, the free functions  $p$ and  $q$ in Eq.~(17)
are chosen so that system (6)--(9) is reduced to a single equation~(14)
with the same functions  $p =  p(t)$ and $q = q(t)$
as a result of the transformation (10),~(17).}

\textcolor{blue}{Thus, this important statement is proved. An arbitrary solution of Eq.~(14) for
arbitrary functions $p = p(t)$ and  $q =  q(t)$ generates an exact solution of
the Navier--Stokes equations \hbox{(1)--(4)}. This solution is described by the function
$F = F(z, t)$ and Eqs.~(5), (10),~(17).}

\textcolor{blue}{\eqnitem 2.
{\sl Case of $m\not=n$\/}.
In this case, the general solution of system (11)--(13) can be obtained as follows.
The functions $m = m(t)$, $k = k(t)$, and $q = q(t)$
are set arbitrarily under the condition $m^2 + k^2\not=1/4$.
The remaining functions included in system (11)--(13) and Eq.~(14)
are calculated sequentially from the formulas}
\begin{equation}
\begin{array}{c}
\ds n=\frac{1-4k^2}{4m},\\[6pt]
A=mq+m'_t,\quad B=nq+n'_t,\quad C=kq+k'_t,\\[6pt]
\ds p=\frac{A'_t-B'_t}{n-m}, \quad
\ds \alpha=AB+C^2+C'_t-\frac12p(1-2k),\\[6pt]
\beta=AB+C^2-C'_t-\ds\frac12p(1+2k),\quad
\gamma=pm+A'_t.
\end{array}
\label{(18)}
\end{equation}
\textcolor{blue}{In this case, the coefficient $p = p(t)$ in Eq.~(14) is determined through the
functions $m = m(t)$, $k = k(t)$, and $q = q(t)$ and their derivatives
(instead of being set arbitrarily as in the case of $m = n$).
An attempt to set $p = p(t)$ directly instead of the function $m$ (or $k$)
results in a nonlinear ordinary differential equation of the second order for
the function $m$ (or $k$) with an arbitrary function $q = q(t)$.}

\textcolor{blue}{We consider how to choose the function $q = q(t)$ so that the identity
$p\equiv 0$ is satisfied. From the expression for~$p$ in Eq.~(18), we have
$A = B + s_0$, where $s_0$ is an arbitrary constant.
From here, taking into account Eqs.~(18) for $A$,~$B$, and~$n$,
we find the function}
\begin{equation}
q=\frac{4s_0m-8(mm'_t+kk'_t)}{4(m^2+k^2)-1}+\frac {m'_t}m\qquad (\hbox{for} \ p=0).
\label{(19)}
\end{equation}

\section{Exact solutions of Eq.~(14) for various $\boldmath p=p(t)$ and
$\boldmath q=q(t)$}

\textcolor{blue}{\eqnitem 1. Functional separable solution:}
$$
F=-a'_t(t)+b(t)[z+a(t)]-\frac{6\nu}{z+a(t)},\quad
q=-4b,\quad p=b'_t+3b^2,
$$
\textcolor{blue}{where $a=a(t)$ and $b=b(t)$ are arbitrary functions.}

\textcolor{blue}{\eqnitem 2. Periodic solutions in the form of the product of functions
of different arguments:}
\begin{equation}
\begin{array}{rl}
F&=a(t)\sin(\sigma z+B),\quad\ds a(t)=C\exp\bl[-\nu \sigma ^2t+\int q(t)\,dt\Br],\\[6pt]
p&=-\sigma ^2a^2(t),\quad \hbox{$q=q(t)$ is an arbitrary function},
\end{array}
\label{(20)}
\end{equation}
\textcolor{blue}{where $B$, $C$, and $\sigma $ are arbitrary constants.
Putting in Eq.~(20) that $q(t)=\nu \sigma ^2+\varphi'_t(t)$, where $\varphi(t)$
is a periodic function, we obtain a solution periodic in both
arguments $z$ and~$t$.}
\medskip

\textcolor{blue}{{\it Example\/}.
Consider the stationary case. In Eqs.~\hbox{(17), (20)}, we put}
$$
\varphi=0,\quad a=-\frac{a_1+a_2}\sigma,\quad q=\nu\sigma^2=2a_1,\quad p=-a^2\sigma^2,\quad
\sigma=(2a_1/\nu)^{\!1/2}.
$$
\textcolor{blue}{As a result, using Eqs.~(5) and (10), we obtain the solution}
$$
V_1=a_1x,\quad V_2=[(a_1+a_2)\cos(\sigma z)-a_1]y,\quad
V_3=-\frac{a_1+a_2}\sigma\sin(\sigma z),
$$
\textcolor{blue}{which describes the three-dimensional flow of a fluid layer between two
flat elastic films (the film position depends on the values of
$z=0$ and $z=2\pi/\sigma$), the surfaces of which are stretched according to the law
$V_1=a_1x$ and $V_2=a_2y$.}

\textcolor{blue}{\eqnitem 3. Generalized separable solutions exponential in~$z$:}
\begin{equation}
F=a(t)e^{-\sigma z}+b(t),\quad p=0,\quad q=\frac {a'_t}a-\sigma b-\sigma ^2\nu,
\label{(21)}
\end{equation}
\textcolor{blue}{where $a=a(t)$ and $b=b(t)$ are arbitrary functions. Choosing $a(t)$ and
$b(t)$ to be periodic functions, we obtain a solution periodic in time.}

\textcolor{blue}{Formulas (20) and (21) together with Eqs.~(5), (10), (17) define two classes of
solutions of the Navier--Stokes equations dependent on several arbitrary functions.}

\textcolor{blue}{\eqnitem 4.
Solution (21) can be represented in the form}
\begin{equation}
F=a_0\exp\Bl[-\sigma z+\sigma ^2\nu t+\int(q+\sigma b)\,dt\Br]+b(t), \quad p=0,
\label{(22)}
\end{equation}
\textcolor{blue}{where $b=b(t)$ and $q=q(t)$ are arbitrary functions and $a_0$ is an arbitrary constant.
Formula (22) defines a new class of exact solutions of the Navier--Stokes
equations with the help of Eqs.~\hbox{(5), (18)}, for $p = 0$ and Eq.~(19).}

\textcolor{blue}{\eqnitem 5. Exact solution in the form of the product of functions of
different arguments:}
$$
F=a(t)(C_1e^{\sigma z}+C_2e^{-\sigma z}),\quad p=4C_1C_2\sigma ^2a^2(t),\quad
q=\frac{a'_t}a-\sigma ^2\nu,
$$
\textcolor{blue}{where $a=a(t)$ is an arbitrary function, $C_1$, $C_2$, and $\sigma $ are arbitrary constants.}

\textcolor{blue}{\eqnitem 6. Monotonic traveling-wave solution:}
$$
F=-6\nu \sigma \tanh[\sigma (z-\lambda t)+B]+\lambda,\quad p=0,\quad q=8\nu \sigma ^2.
$$

\textcolor{blue}{\eqnitem 7. Unbounded periodic traveling-wave solution:}
$$
F=6\nu \sigma \tan[\sigma (z-\lambda t)+B]+\lambda,\quad p=0,\quad q=-8\nu \sigma ^2.
$$

\textcolor{blue}{\eqnitem 8. Functional separable solution:}
$$
F=\frac{a(t)}{\lambda(t)}\exp[-\lambda(t)z]+b(t)+c(t)z,
$$
\textcolor{blue}{where the functions  $a=a(t)$, $b=b(t)$, $c=c(t)$, and $\lambda=\lambda(t)$
satisfy the system of ordinary differential equations}
$$
\lambda'_t=-c\lambda, \ \quad
a'_t=(\nu \lambda^2+q+2c+b\lambda)a, \ \quad
c'_t=c^2+qc+p.
$$
\textcolor{blue}{Here, three of the six functions  $a(t)$, $b(t)$, $c(t)$, $\lambda(t)$,
$p(t)$, and $q(t)$ can be set arbitrarily.}

\textcolor{blue}{\eqnitem 9. Functional separable solution:}
\begin{equation}
F=\omega(t)z+\frac{\xi(t)}{\theta(t)}\sin[\theta(t)z+a],
\label{(23)}
\end{equation}
\textcolor{blue}{where $a$ is an arbitrary constant, and the functions  $\omega=\omega(t)$, $\xi=\xi(t)$, and
$\theta=\theta(t)$  are described by the system of ordinary differential equations}
\begin{equation}
\theta'_t=-\omega\theta, \ \quad
\omega'_t=\omega^2+q(t)\omega+p(t)+\xi^2, \ \quad
\xi'_t=[2\omega-\nu\theta^2+q(t)]\xi.
\label{(24)}
\end{equation}
\textcolor{blue}{In this system, it is possible to treat the functions
$\theta(t)$ and $\xi(t)$ as arbitrary, whereas the functions $\omega(t)$, $p(t)$,
and $q(t)$ are elementarily determined (without quadratures).
A periodic solution~(23) corresponds to periodic functions
$\theta(t)$ and $\xi(t)$.}

\textcolor{blue}{\eqnitem 10. Functional separable solution:}
\begin{equation}
F=\omega(t)z+\frac{\xi(t)}{\theta(t)}\Bl[C_1e^{\theta(t)z}+C_2e^{-\theta(t)z}\Br],
\label{(25)}
\end{equation}
\textcolor{blue}{where $C_1$ and $C_2$ are arbitrary constants, and the functions $\omega=\omega(t)$, $\xi=\xi(t)$, and
$\theta=\theta(t)$ are described by the system of ordinary differential equations}
\begin{equation}
\theta'_t=-\omega\theta, \ \quad
\omega'_t=\omega^2+q(t)\omega+p(t)-4C_1C_2\xi^2, \ \quad
\xi'_t=[2\omega+\nu\theta^2+q(t)]\xi.
\label{(26)}
\end{equation}

\textcolor{blue}{{\it Remark\/}.
See also [8, 13, 15] for exact solutions of Eq.~(14) with $q = 0$.}

\section{Reduction of system (6)--(9) to two equations}

\textcolor{blue}{We describe two cases of reducing system (6)--(9)
to a single isolated nonlinear equation for the longitudinal velocity $F$
and a second equation for determining a new auxiliary function.}

\textcolor{blue}{\eqnitem 1. {\sl First case\/.} By letting}
\begin{equation}
u=a^2G,\quad v=-b^2G,\quad w=abG,\quad
\alpha=\beta,\quad \gamma=0,
\label{(27)}
\end{equation}
\textcolor{blue}{where $a$ and $b$ are arbitrary constants,
we reduce system (6)--(9) to an isolated equation for the longitudinal velocity~$F$
and an additional equation for the function~$G=G(z,t)$:}
\begin{eqnarray}
\pdd Ftz+F\pdd Fzz-\frac12\Bl(\pd Fz\Br)^2&=&-2\alpha+
\nu \frac{\partial^3F}{\partial z^3},\label{(28)}\\
\pd Gt+F\pd Gz-G\pd Fz&=&\nu\pdd Gzz.
\end{eqnarray}

\textcolor{blue}{\eqnitem 2.
{\sl Second case\/.} In Eqs.~\hbox{(6)--(9)}, let}
\begin{equation}
\begin{array}{rl}
u&=\ds\frac12\sin\varphi\Bl(\pd Fz+q\Br)+a^2\Theta,\\[6pt]
v&=\ds\frac12\sin\varphi\Bl(\pd Fz+q\Br)-b^2\Theta,\\[6pt]
w&=\ds\frac12\cos\varphi\Bl(\pd Fz+q\Br)+ab\Theta,\\[6pt]
\alpha&=\ds\frac14q^2-\frac12p(1-\cos\varphi)+\frac12q'_t\cos\varphi,\\[6pt]
\beta&=\ds\frac14q^2-\frac12p(1+\cos\varphi)-\frac12q'_t\cos\varphi,\\[6pt]
\gamma&=\ds\frac12p\sin\varphi+\frac12q'_t\sin\varphi,
\end{array}
\label{(30)}
\end{equation}
\textcolor{blue}{where
$p=p(t)$ and $q=q(t)$ are arbitrary functions, and $a$ and $b$ are arbitrary constants,
$\Theta=\Theta(z,t)$
is an unknown function, and $\varphi$ is the constant determined from the
transcendental equation}
\begin{equation}
(a^2-b^2)\sin\varphi+2ab\cos\varphi=0.
\label{(31)}
\end{equation}
\textcolor{blue}{As a result, system (6)--(9) is reduced to two equations:}
\begin{eqnarray}
\pdd Ftz+F\pdd Fzz-\Bl(\pd Fz\Br)^2&=&
\nu \frac{\partial^3F}{\partial z^3}+q\pd Fz+p,\label{(32)}\\
\pd \Theta t+F\pd \Theta z-\Theta \pd Fz&=&\nu\pdd \Theta zz.
\end{eqnarray}
\textcolor{blue}{The nonlinear equation~(32) for $F$ coincides with Eq.~(14)
and can be treated independently (some of its exact solutions were described
previously), and Eq.~(33) is linear with respect to the desired
function~$\Theta$.}

\textcolor{blue}{For stationary solutions of Eqs.~(32) and (28) (for constant $p$, $q$, and $\alpha$),
the nonstationary equations~(33) and (29) are linear separable equations,
whose solutions can be obtained using the Laplace transform in time.}

\textcolor{blue}{Equation (32) (and Eq.~(28)) admits an obvious degenerate solution $F = a(t)z + b(t)$;
in this case, the corresponding Eq.~(33) (and Eq.~(29)) can be reduced to the
linear heat equation.}

\textcolor{blue}{System (28), (29) for an arbitrary function $\alpha=\alpha(t)$ has a solution in the form}
\begin{eqnarray*}
F&=&az^2+b(t)z+\frac 1{4a}[b^2(t)-2b'_t(t)-4\alpha(t)],\\
w&=&A(t)z^2+B(t)z+C(t),
\end{eqnarray*}
\textcolor{blue}{where $a$ is an arbitrary constant ($a\not=0$), $b(t)$ is an arbitrary function,
and the functions  $A=A(t)$, $B=B(t)$, and $C=C(t)$
are described by the system of ordinary differential equations,
which is not presented here.}

\section{Interpretation and classification of the flows under consideration}

\textcolor{blue}{Arbitrary fluid flows having two symmetry planes admit a representation of the
type of Eq.~(5) in the vicinity of the line of intersection of these planes
(in the adopted notation, the planes intersect in the $z$ axis).
Such flows include the axisymmetric flows, combinations of axisymmetric flows
with rotation around of the $z$ axis (in particular, the von Karman type flows),
plane flows symmetric with respect to a straight line, flows in rectilinear
impenetrable and porous pipes with elliptic and rectangular cross sections,
fluid jets flowing from orifices of elliptic and rectangular shapes, etc.\ (see also [10,~11]).}

\textcolor{blue}{It is convenient to treat the axial flows described by
Eqs.~(5) as a nonlinear superposition of a translatory (nonuniform) flow along
the $z$ axis and a linear shear flow of a special type. In the vicinity of
the point $z = z_0$ lying on the axis, the components of fluid velocity taking
into account Eq.~(5) can be represented as}
\begin{equation}
\begin{array}{c}
V_k=F\delta_{k3}+ G_{km}X_m;\\[6pt]
G_{11}=-\frac12F_z+w,\quad G_{12}=v,\quad
G_{21}=u,\quad G_{22}=-\frac12F_z-w,\\[6pt]
G_{13}=G_{23}=G_{31}=G_{32}=0,\quad G_{33}=F_z;\\[6pt]
X_1=x,\quad X_2=y,\quad X_3=z-z_0.
\end{array}
\label{(34)}
\end{equation}
\textcolor{blue}{Here $k,\,m=1,\,2,\,3$;
the $G_{km}$ are shear matrix components;
the summation is assumed over the repeated subscript~$m$;
$\delta_{km}$ is the Kronecker delta; and $F_z$ is the partial derivative with respect to~$z$.
All values in Eq.~(34) are taken for $z=z_0$.
The vanishing of the sum $G_{11}+G_{22}+G_{33}=0$
of diagonal elements is a consequence of fluid incompressibility.}

\textcolor{blue}{An arbitrary matrix $\|G_{km}\|$ can be represented in the form of the sum of
a symmetric and an asymmetric matrix}
\begin{equation}
\begin{array}{c}
\|G_{km}\|=\|E_{km}\|+\|\Omega_{km}\|,\\[6pt]
E_{km}=E_{mk}=\frac 12 (G_{km}+G_{mk}),\quad
  \Omega_{km}=-\Omega_{mk}=\frac 12 (G_{km}-G_{mk}).
\end{array}
\label{(35)}
\end{equation}

\textcolor{blue}{In turn, the symmetric matrix  $\|E_{km}\|$
(in this case, it can be identified with the strain-rate tensor) can
be reduced to a diagonal form with diagonal elements
$E_1$, $E_2$, and~$E_3$, which are roots of the cubic equation
$\det\|E_{km}-\lambda\delta_{km}\|=0$ for $\lambda$,
by appropriately rotating the system of coordinates.}

\textcolor{blue}{For this flow (34), the diagonal elements determining the intensity of the tension
(compression) motion along the respective axes are calculated from the formulas}
\begin{equation}
E_{1,2}=-\frac 12F_z\pm\frac12\sqrt{4w^2+(u+v)^2},\quad E_3=F_z.
\label{(36)}
\end{equation}

\textcolor{blue}{The splitting of the shear coefficient matrix $\|G_{km}\|$
into symmetric and asymmetric parts (35) corresponds to the representation
of the velocity field of the linear shear flow of the fluid as a superposition
of a linear deformational flow with the tension coefficients
$E_1$, $E_2$, and~$E_3$
along the principal axes and rotations of the fluid as a solid body with the
angular velocity
$\vec\omega =(\Omega_{32},\Omega_{13},\Omega_{21})$.}

\textcolor{blue}{For this flow (34), we have
$\Omega_{32}=\Omega_{13}=0$ and the fluid rotates around the $z$
axis with the angular velocity}
\begin{equation}
\Omega_{21}=\frac12(u-v).
\label{(37)}
\end{equation}

\textcolor{blue}{It is easy to show that Eqs.~(36) and (37) remain valid for an arbitrary point
$(x_0,y_0,z_0)$ of the flow (5) under consideration.}

\textcolor{blue}{The analysis of Eqs.~\hbox{(36), (37)} enables us to single out certain characteristic
types of flows indicated in the classification table.}
\bigskip

\begin{table}
\renewcommand{\arraystretch}{1.5}
\footnotesize
\caption{\footnotesize Classification of axial flows described by Eqs.~(5)}

\begin{center}
\begin{tabular}{|c|c|c|}
\hline
  Type of flow
  & Desired functions
  & Functions included in pressure
 \\ \hline
  Axisymmetric
  & $u=v=w=0$
  & $\alpha=\beta$, \ $\gamma=0$
 \\ \hline
   \vrule height 1.8em depth 1.2em width 0pt$\vcenter{
    \hbox{Combination of axisymmetric flow\strut}
    \hbox{and rotation around of the $z$ axis\strut}}$
  & $w=0$, \ $v=-u$
  & $\alpha=\beta$, \ $\gamma=0$
 \\ \hline
   Pure deformational (without rotation)
  & $u=v$
  & $\alpha,\beta,\gamma$ are arbitrary functions
 \\ \hline
   General axial
  & $u\not=v$
  & $\alpha,\beta,\gamma$ are arbitrary functions
 \\ \hline
\end{tabular}
\end{center}
\end{table}

\section{Some generalizations}

\textcolor{blue}{Let $V_1(x,y,z,t)$, $V_2(x,y,z,t)$, $V_3(x,y,z,t)$, and $P(x,y,z,t)$
be a certain solution of Navier--Stokes equations~\hbox{(1)--(4)}. Then the set of functions}
\begin{equation}
\begin{array}{rl}
\bar V_1&=V_1(x-x_0,y-y_0,z-z_0,t)+x_0',\\
\bar V_2&=V_2(x-x_0,y-y_0,z-z_0,t)+y_0',\\
\bar V_3&=V_3(x-x_0,y-y_0,z-z_0,t)+z_0',\\
\bar P&=P(x-x_0,y-y_0,z-z_0,t)-\rho(x_0''x+y_0''y+z_0''z),
\end{array}
\label{(38)}
\end{equation}
\textcolor{blue}{where $x_0=x_0(t)$, $y_0=y_0(t)$, and $z_0=z_0(t)$ are arbitrary functions
(primes denote the derivatives with respect to $t$),
also gives the solution of Eqs.~\hbox{(1)--(4)} [4, 15].
The combination of Eqs.~(5) and (38) for $z_0 = 0$ determines an exact solution
of the Navier--Stokes equations, which can be treated as the generalized axial
flow with the $z$ axis moving along the plane $x$, $y$,
according to the law $x = x_0(t)$, $y = y_0(t)$. The indicated solution can be used
for the mathematical simulation of destructive atmospheric phenomena such as
waterspouts and tornados.}

\section*{Acknowledgments}

\textcolor{blue}{The authors thank A.~N.~Osiptsov for useful remarks.}

\textcolor{blue}{The work was carried out under partial financial support of the Russian
Foundation for Basic Research (grants  No.~01-08-00553, No.~08-08-00530, and \hbox{No.~07-01-96003-r$_{-}$ural$_{-}$a}).}

\end{document}